\documentclass[12pt]{article}
\usepackage{graphicx}
\begin{document}
\begin{center}

\vspace*{3cm}

\textbf{\LARGE The time evolution\\ \smallskip
of the Bohmian Pilot Wave}

\bigskip
\bigskip
{\Large M. Kozlowski, J. Marciak-Kozlowska}

\bigskip

Institute of Electron Technology
Al. Lotnik\'{o}w 32/46, 02-668 Warsaw, Poland

\end{center}

\vspace{2cm}
\begin{abstract}
In this paper the Newton-Schr\"{o}dinger-Bohm equation is solved for particles with $m>M_P$. It is shown that the Bohmian pilot wave for particles with $m>M_P$ oscillates with frequency $\omega=\tau_P^{-1}$, where $\tau_P$ is the Planck time.
\end{abstract}

\newpage
\section{The Newton-Schr\"{o}dinger-Bohm equation}
D.~Bohm presented the pilot wave theory in 1952 and de  Broglie had presented a similar theory in the mid 1920's. It was rejected in 1950's and the rejection had nothing to do with de Broglie and Bohm later works.

There is always the possibility that the pilot wave has a primitive, mind like property. That's how Bohm described it. We can say that all the particles in the Universe end even Universe have their own pilot waves, their own information. Then the consciousness for example is the very complicated receiver of the surrounding pilot wave fields.

In our paper, http://lanl.arxiv.org/quant-ph/0402069/ a study of the Newton-Schr\"{o}dinger-Bohm (NSB) equation for the pilot wave was developed:
\begin{equation}
i\hbar\frac{\partial \Psi}{\partial t}=-\frac{\hbar^2}{2m}\nabla^2\Psi+V\Psi-\frac{\hbar^2}{2M_P}\nabla^2\Psi+\frac{\hbar^2}
{2M_P}\left(\nabla^2\Psi-\frac{1}{c^2}\frac{\partial^2\Psi}{\partial t^2}\right).
\label{eq1}
\end{equation} 
In Eq.~(\ref{eq1}) $m$ is the mass of the quantum particle and $M_P$ is the Planck mass $(M_P\approx10^{-5}~{\rm g})$.

For elementary particles with mass $m<<M_P$ we obtain from Eq.~(\ref{eq1})
\begin{equation}
i\hbar\frac{\partial \Psi}{\partial t}=-\frac{\hbar^2}{2m}\nabla^2 \Psi+V\Psi+\frac{\hbar^2}{2M_P}\left(\nabla^2\Psi-\frac{1}{c^2}\frac{\partial^2\Psi}{\partial t^2}\right)\label{eq2}
\end{equation}
and for macroscopic particles with $m>>M_P$ equation~({\ref{eq1}) has the form:
\begin{equation}
i\hbar\frac{\partial \Psi}{\partial t}=-\frac{\hbar^2}{2M_P}\nabla^2\Psi+\frac{\hbar^2}{2M_P}\left(\nabla^2\Psi-\frac{1}{c^2}
\frac{\partial^2\Psi}{\partial t^2}\right)+V\Psi\label{eq3}
\end{equation}
or
$$
i\hbar\frac{\partial\Psi}{\partial t}=-\frac{\hbar^2}{2M_Pc^2}\frac{\partial^2\Psi}{\partial t^2}+V\Psi
$$
and is dependent of $m$.
\section{Discussion of the results}
In the following we will discuss the pilot wave time evolution for the macroscopic particles, i.e. for particles with $m>>M_P$.

For $V=$~const we seek the solution of Eq.~(\ref{eq3}) in the form
\begin{equation}
\Psi=e^{\gamma t}.\label{eq4}
\end{equation}
After substitution formula~(\ref{eq4}) to Eq.~(\ref{eq3}) one's obtains
\begin{equation}
M_P\gamma^2+\frac{2M_P^2c^2}{\hbar}\gamma-\frac{2M_P^2c^2}{\hbar^2}V=0\label{eq5}
\end{equation}
with the solution
\begin{equation}
\gamma_{1,2}=-\frac{iM_Pc^2}{\hbar}\pm\frac{M_Pc^2}{\hbar}\sqrt{-1+\frac{2V}{M_Pc^2}}.\label{eq6}
\end{equation}
For a free particle, $V=0$ we obtain:
\begin{equation}
\gamma_{1,2}=\left\{\begin{array}{c}
  0, \\
  -\frac{2M_Pc^2}{\hbar}i. 
\end{array}\right.\label{eq7}
\end{equation}
According to formulae~(\ref{eq4}) and (\ref{eq7}) equation~(\ref{eq3}) has the solution
\begin{equation}
\Psi(t)=A+Be^{-\frac{2M_Pc^2i}{\hbar}t}.\label{eq8}
\end{equation}
For $t=0$ we put $\Psi(0)=0$, then
$$
\Psi(t)=A\left(1-e^{-\frac{2ti}{\tau_P}}\right),
$$
where $\tau_P=$ Planck time
\begin{equation}
\tau=\frac{\hbar}{M_Pc^2}.\label{eq9}
\end{equation}
In Fig.~\ref{fig1} $\Re\Psi(t)$ is presented for $t\in(0, 100\tau)$ and $t\in(0,1000\tau)$ respectively. From Fig.~\ref{fig1} one's can conclude that the free particle in reality is jittering with frequency $\omega=\tau^{-1}$ and quantum energy $E=\hbar\omega$,
\begin{equation}
E=\hbar\omega=10^{19}{\rm GeV}\label{eq10}
\end{equation}
and period $T$
$$
T=10^{-43}{\rm s}.
$$
Considering that the contemporary limiting time resolution is of the order $10^{-18}$~s (attosecond laser pulses) the jittering of the free macroscopic particle can not be observed.
\section{Conclusion}
The NSB equation offers the new insight into the realm of the subquantum structure of the space-time. It occurs that free macroscopic particles, with mass $m>M_P$ are embedded in the field of the oscillations with the period of $T\approx10^{-43}$~s and energy $E\approx10^{19}$~GeV.
\begin{figure}[p]
\centering\includegraphics[scale=0.6]{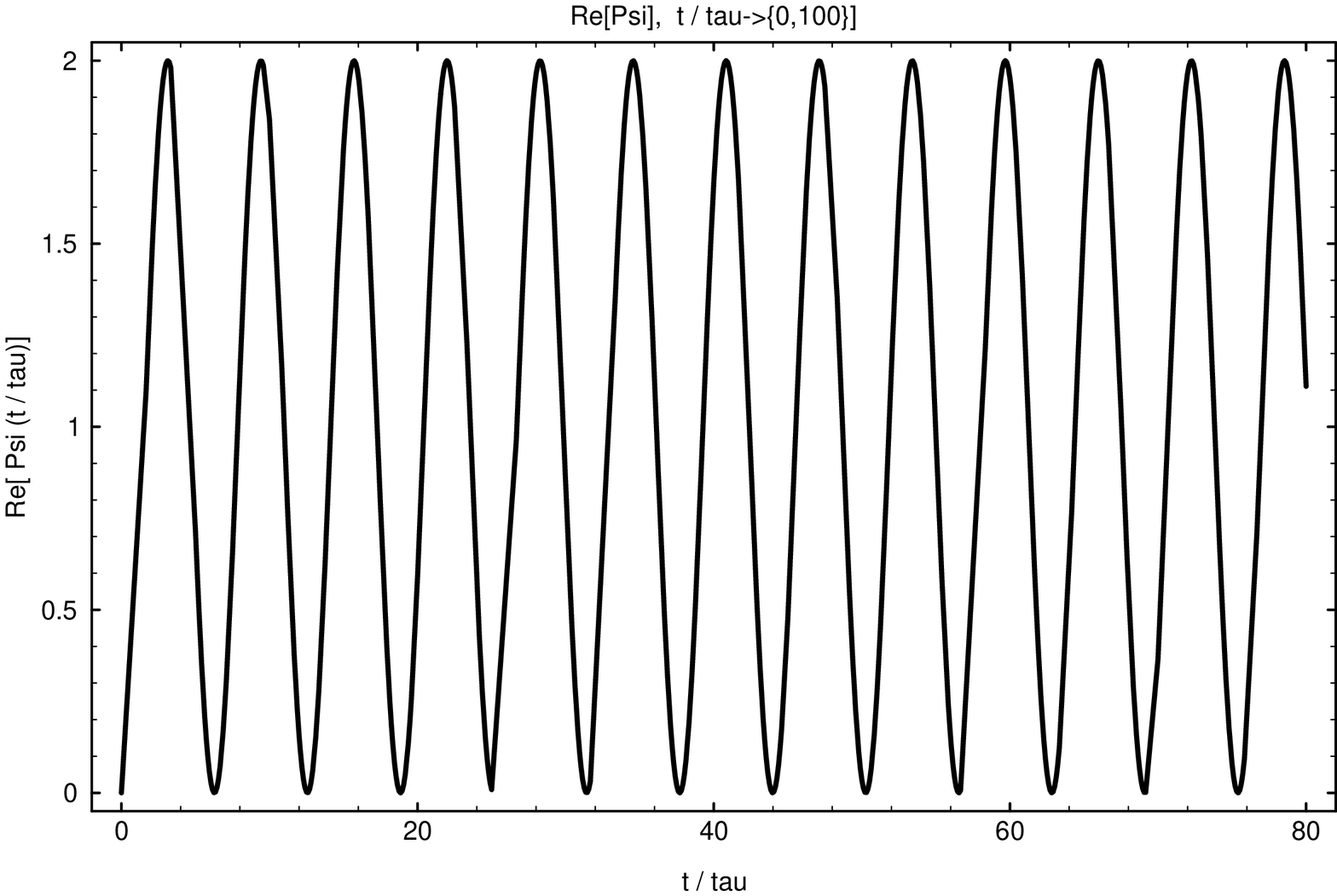}

\centering\includegraphics[scale=0.6]{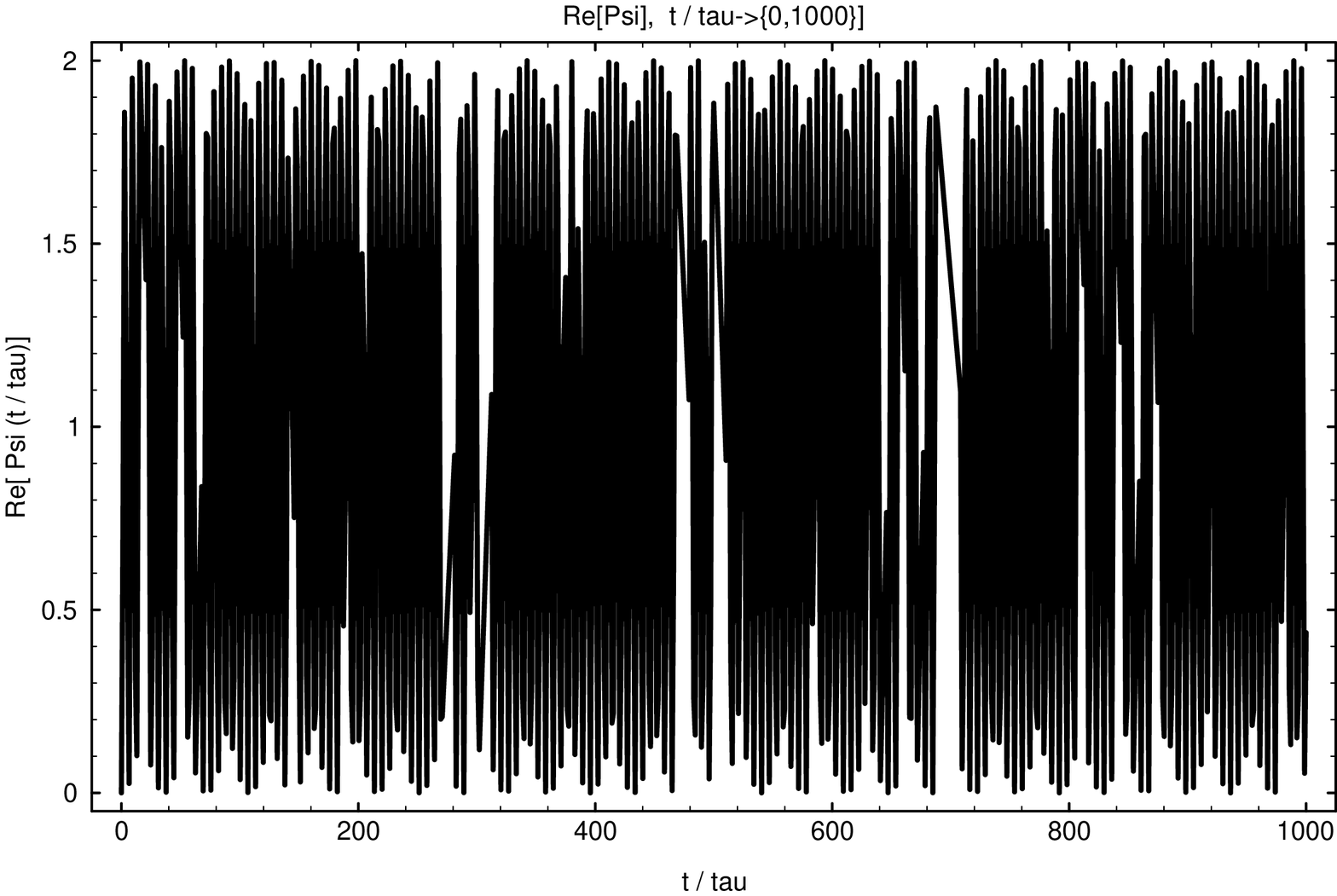}
\caption{}\label{fig1}
\end{figure}

\end{document}